%% file: main.tex
\documentclass[conference]{IEEEtran}
\usepackage{cite}
\usepackage{amsmath,amssymb,amsfonts}
\usepackage{algorithmic}
\usepackage{graphicx}
\usepackage{textcomp}
\usepackage{xcolor}
\def\BibTeX{{\rm B\kern-.05em{\sc i\kern-.025em b}\kern-.08em
    T\kern-.1667em\lower.7ex\hbox{E}\kern-.125emX}}
    
\input{preamble} 
    
\begin{document}

\title{Test Scenario Generation for Context-Oriented Programs}

\author{
\IEEEauthorblockN{Pierre Martou}
\IEEEauthorblockA{\textit{ICTEAM, UCLouvain} \\
Louvain-la-Neuve, Belgium \\
pierre.martou@uclouvain.be}
\and
\IEEEauthorblockN{Kim Mens}
\IEEEauthorblockA{\textit{ICTEAM, UCLouvain} \\
Louvain-la-Neuve, Belgium \\
kim.mens@uclouvain.be}
\and
\IEEEauthorblockN{Benoît Duhoux}
\IEEEauthorblockA{\textit{ICTEAM, UCLouvain} \\
Louvain-la-Neuve, Belgium \\
benoit.duhoux@uclouvain.be}
\and
\IEEEauthorblockN{Axel Legay}
\IEEEauthorblockA{\textit{ICTEAM, UCLouvain} \\
Louvain-la-Neuve, Belgium \\
axel.legay@uclouvain.be}
}

\maketitle

\input{Abstract}


\input{Introduction}
\input{FBCOP}

\input{Testing}
\input{CIT}

\input{Rearrangement}
\input{Incremental}
\input{Future-work}
\input{Acknowledgements}

\bibliographystyle{ieeetr}
\bibliography{biblio.bib}

%\appendix

\end{document}

%% file: preamble.tex
\usepackage[utf8]{inputenc}		% to allow for French accents
\usepackage{xcolor,colortbl}
\usepackage{ifdraft}
\usepackage{multirow}
\usepackage{graphics}      % for EPS, load graphicx instead 
\usepackage{mathtools}

\usepackage{mathptmx}
\definecolor{author}{rgb}{.5, .5, .5}
\definecolor{comment}{rgb}{.1, .0, .9}
\definecolor{note}{rgb}{.7, .4, .0}
\definecolor{idea}{rgb}{.1, .7, .0}
\definecolor{missing}{rgb}{.9, .1, .0}

\usepackage[ruled,vlined,linesnumbered]{algorithm2e}
\include{pythonlisting}
\usepackage{tabularx}
\usepackage{makecell}

%% file: Abstract.tex
\begin{abstract}
Their highly adaptive nature and the combinatorial explosion of possible configurations makes testing context-oriented programs hard.
We propose a methodology to automate the generation of test scenarios for developers of feature-based context-oriented programs.
By using combinatorial interaction testing we generate a covering array from which a small but representative set of test scenarios can be inferred.
By taking advantage of the explicit separation of contexts and features in such context-oriented programs, we can further rearrange the generated test scenarios to minimise the reconfiguration cost between subsequent scenarios.
Finally, we explore how a previously generated test suite can be adapted incrementally when the system evolves to a new version.
By validating these algorithms on a small use case, our initial
results show that the proposed test generation approach is efficient and beneficial to developers to test and improve the design of context-oriented programs.
\end{abstract}

%\begin{CCSXML}
%<ccs2012>
%<concept>
%<concept_id>10011007.10011074.10011099.10011102.10011103</concept_id>
%<concept_desc>Software and its engineering~Software testing and debugging</concept_desc>
%<concept_significance>500</concept_significance>
%</concept>
%<concept>
%<concept_id>10011007.10011074.10011099</concept_id>
%<concept_desc>Software and its engineering~Software verification and %validation</concept_desc>
%<concept_significance>500</concept_significance>
%</concept>
%</ccs2012>
%\end{CCSXML}

%\ccsdesc[500]{Software and its engineering~Software testing and debugging}
%\ccsdesc[500]{Software and its engineering~Software verification and validation}

%new
%\keywords{context-oriented programming, combinatorial interaction testing, covering array, test scenario generation, satisfiability (SAT) checking}


%% file: Introduction.tex
\section{Introduction}
\textit{Context-Oriented Programming} (COP) languages help developers build context-aware applications that, based on contextual information sensed from the surrounding environment, can adapt their behaviour dynamically. A wide range of COP languages have been proposed to ease the design and implementation of such applications~\cite{Costanza05ContextL, Hirschfeld08ContextS, Gonzalez11SubjectiveC, Salvaneschi11Javactx}.
%~\cite{Appeltauer08ContextJ, Gonzalez08Ambient, Ghezzi10Programming, Costanza08FDCOP, Aotani11Featherweight, Lincke11ContextJS, Salvaneschi12Contexterlang, Gonzalez13ContextTraits}
%
\textit{Feature-Based Context-Oriented Programming} (FBCOP) \cite{Duhoux19ImplFBCOPL, Duhoux19SATToSE} is a particular class of such languages that takes inspiration from context-oriented programming~\cite{Hirschfeld08COP, Salvaneschi12COPSurvey, Cardozo11FOPCOP, Cardozo14FeaturesOnDemand}, Feature Modelling (FM)~\cite{Kang90FODA} and Dynamic Software Product Lines (DSPL)~\cite{Hartmann08FeaturesContexts, Capilla14ContextVariability, Mens17ModelingManaging}.
% ORIGINAL REFS: Hartmann08FeaturesContexts, Jaroucheh2010,  Capilla14ContextVariability, Murguzur2014, Mens17ModelingManaging

Whereas COP and FBCOP languages claim to provide better support for dynamic adaptability than traditional programming languages, traditional testing techniques meet their limits when applied to context-oriented applications. Dedicated techniques to generate test scenarios are needed to cope with the high variability of the environment (context) and how the application can change its behaviour (features) to situations in that environment. Such a test scenario defines a specific context in which the application operates.

This paper pursues the following three research questions, each of which coincides with one of the contributions of this paper.
\begin{description}
\item[RQ1] How to generate a pertinent yet tractable set of test scenarios for a given FBCOP application?
\item[RQ2] %How to group these test scenarios so that they can be presented more concisely?
%\authorcomment[comment]{BD-PM}{Question reformulée; plus générale et intéressante pour un tester qui ne connait rien au FBCOP. Effort relié au creation cost en 3.2.}
%How to decrease the effort of using this test suite ?
How to minimise the effort of creating these generated test scenarios?
\item[RQ3] How to incrementally adapt a previously generated set of scenarios upon evolution of the application to a new version?
\end{description}

To address RQ1 we explore the use of (pairwise) \emph{Combinatorial Interaction Testing} (CIT) to generate a small yet representative set of test scenarios that covers all valid combinations of pairs of contexts or features. We reduce the problem of generating test scenarios for FBCOP programs to one of computing the \emph{covering array} of a highly reconfigurable system in the presence of constraints, granting direct access to an efficient greedy SAT-solving algorithm \cite{cohen2008constructing}.

Whereas the algorithm explored to address RQ1 tries to minimise the number of tests needed to cover all interactions between pairs of contexts or features, the ordering of the different scenarios in the generated test suite may not be optimal.
A software tester may desire the scenarios to be automatically ordered in such a way
that a minimal amount of reconfiguration of the context-aware program to be tested is required between each test scenario.
Such a reconfiguration requires adapting or simulating the program via the activation and deactivation of certain contexts. Since FBCOP programs are modelled in terms of separate context and feature models, we can focus on the context model alone to define a \textit{reconfiguration cost} which is the number of context activation switches needed to go from one test configuration to another. Our answer to RQ2 is then a greedy algorithm to reorder test scenarios in the test suite so as to minimize this cost, effectively reducing the effort for a software tester to implement these tests.
% Using a distance function between contexts, it starts by identifying the closest distance context to a default state then incrementally lists closest distance contexts and related test scenarios until the whole suite has been explored. 

% Previous paragraph on second contribution: As a second contribution, we explore how test suites produced by the above mentioned algorithm can be exploited in practice. One main challenge in applying test suites for context-aware programs is that each test may require to reconfigure the system depending on the context it refers to. In practice, this reorganisation calls for adapting the system via the activation and deactivation of contexts. The cost of such an operation increases with the number of such (de)activations. In order to minimize such context switches, we propose a greedy algorithm to polish and rearrange test scenarios. The algorithm is based on a distance function between contexts. It starts by identifying the most liberal context, and then incrementally lists closest distance contexts and related test scenarios until the whole suite has been explored. 

Finally, regarding RQ3, we want to avoid having to regenerate completely the test scenarios when new contexts or features
appear as the program evolves to a new version. 
Our approach consists of updating test scenarios that were already generated for the previous version of the evolved program. Hoping that it already contains many of the new pairs that should be covered, Cohen's algorithm \cite{cohen2008constructing} is then fed %initialized
with this candidate test suite and produces a small number of test scenarios covering the new pairs that are not yet covered. We propose two different update strategies and show that we can drastically reduce the cost of obtaining an up-to-date test suite by almost a factor 4 on our case study.

% Previous paragraph: A final challenge with test suite production and usage is to avoid recomputing the entire suite when new functionalities are added to the system, as well as to reduce the number of test cases that must be added to cover new configurations. In our FBCOP setting, addition of these functionalities may take the form of adding new contexts and features. Such an addition affects the number of compatible pairs that should be tested. A third contribution of this paper is therefore to explore how the above algorithms can be extended to support such incremental additions. Our approach consists of updating tests that were already produced for the non-incremented system. Cohen's algorithm \cite{cohen2008constructing} is then initialized with this candidate test suite, hoping that it already contains many of the new pairs that should be covered. Our first update strategy is to rely on the incremental process of SAT-solvers. This will automatically generate a valid configuration that includes the new features. On the other hand, since the solver only produces one new configuration, the disparity in combinations between old and new functionalities is likely to be limited. Our second update strategy consists in computing a covering array for the newly added features, and then combine it with the existing test suite. As opposed to the incremental SAT-solver approach, this second approach may propose incomparable configurations.
%However, by using a combined strategy, we show that coverage can be improved, hence reducing the final size of the new test suite.

We implemented each of these three contributions and applied them on a small exploratory case study.
%Each of these three contributions were implemented in a prototype which we applied to a small exploratory case study.
Our initial experiments confirm both the feasibility and efficiency of our approach, effectively addressing each of our three research questions.
%providing possible and sensible answers to the three RQs.
The three contributions combined provide a lightweight methodology and set of algorithms for creating test suites easily and efficiently while developing (feature-based) context-oriented programs.

%The approach goes beyond traditional combinatorial interaction testing by taking into account the clear separation between contexts and features in FBCOP, by improving the efficiency and presenting the results in more polished ways. Furthermore, the side-products of the test-suite generation can provide useful information that can help developers of such systems improve the design of their systems.

The remainder of this paper is structured as follows.
Section \ref{sec:fbcop} introduces the FBCOP approach on top of which we built our test scenario generation approach, as well as the case study used as running example and validation. Section \ref{sec:case-testing} revisits our three research questions and establishes our goals. Section \ref{sec:cit} through \ref{sec:incremental-testing} then explain and validate each of our three contributions in detail. Section \ref{sec:conclusion} concludes the paper and presents interesting avenues of future work.

%% file: FBCOP.tex
%\section{Background}
% Background – part 1

\section{Feature-based context-oriented programming}
\label{sec:fbcop}

Unlike other COP languages, FBCOP makes an explicit distinction between how contexts (reifying particular situations sensed from the surrounding environment) and features (behavioural variations specific to certain contexts) are modelled and handled~\cite{Duhoux19ImplFBCOPL, Duhoux19SATToSE}.

Before introducing FBCOP, we introduce the running example used throughout this paper and the notion of feature diagrams.
Then we exemplify how FBCOP can be used to write programs that adapt their behaviour upon context changes.

\subsection{Smart messaging system}
\label{sec:case}
As a running example of a FBCOP program, we implemented a prototype of a \emph{smart messaging system}.
The system allows users to exchange messages and is smart in the sense that it can adapt or refine its behaviour depending on some contextual situations. 
The system's main functionality consists of \emph{Sending} or \emph{Receiving} messages to and from other users.
By default, we assume the messages are of type \emph{Text} though richer messages (\emph{Vocal} or \emph{Photo}) can be exchanged depending on the status of the internet \emph{Connection} or the \emph{Peripheral} cards (audio or video) installed on the \emph{Device} running the program. 

Conversations can take place between a \emph{Group} of \emph{Friend}s.
Each user is identified by its unique \emph{FriendName}.
Depending on the users' \emph{Age} (\emph{Teen} or \emph{Adult}), they can complete their user profile with a textual \emph{Description} or \emph{ProfilePicture} (respectively). 
How new friends can be added to a conversation also depends on the users' \emph{Age}.
%Conversations can also be carried out between a \emph{Group} of friends, to which more friends can be added during conversation time. The technique to add new friends is also adapted  depending on the user age.

Whenever users receive a new message, a \emph{Notification} may be given via a sound \emph{Alarm} or their device's \emph{Vibration} mode. What notification mode is used depends on the type of \emph{Device}, the \emph{User Availability} and the ambient \emph{Noise} level.
% However the notification system can be muted when the ambient \emph{Noise} level is \emph{Quiet}, so as not to disturb the user, or when the user is \emph{Occupied}.
%Finally 
The \emph{Device} type also affects other features of the smart messaging system, such as adapting the layout that displays the information or adding a virtual keyboard.

%A simplified version of the context, mapping and feature model of this system is depicted in Figure~\ref{fig:casestudy-models}.

\begin{figure*}[!ht]
  \centering
  \includegraphics[draft=false, width=\linewidth]{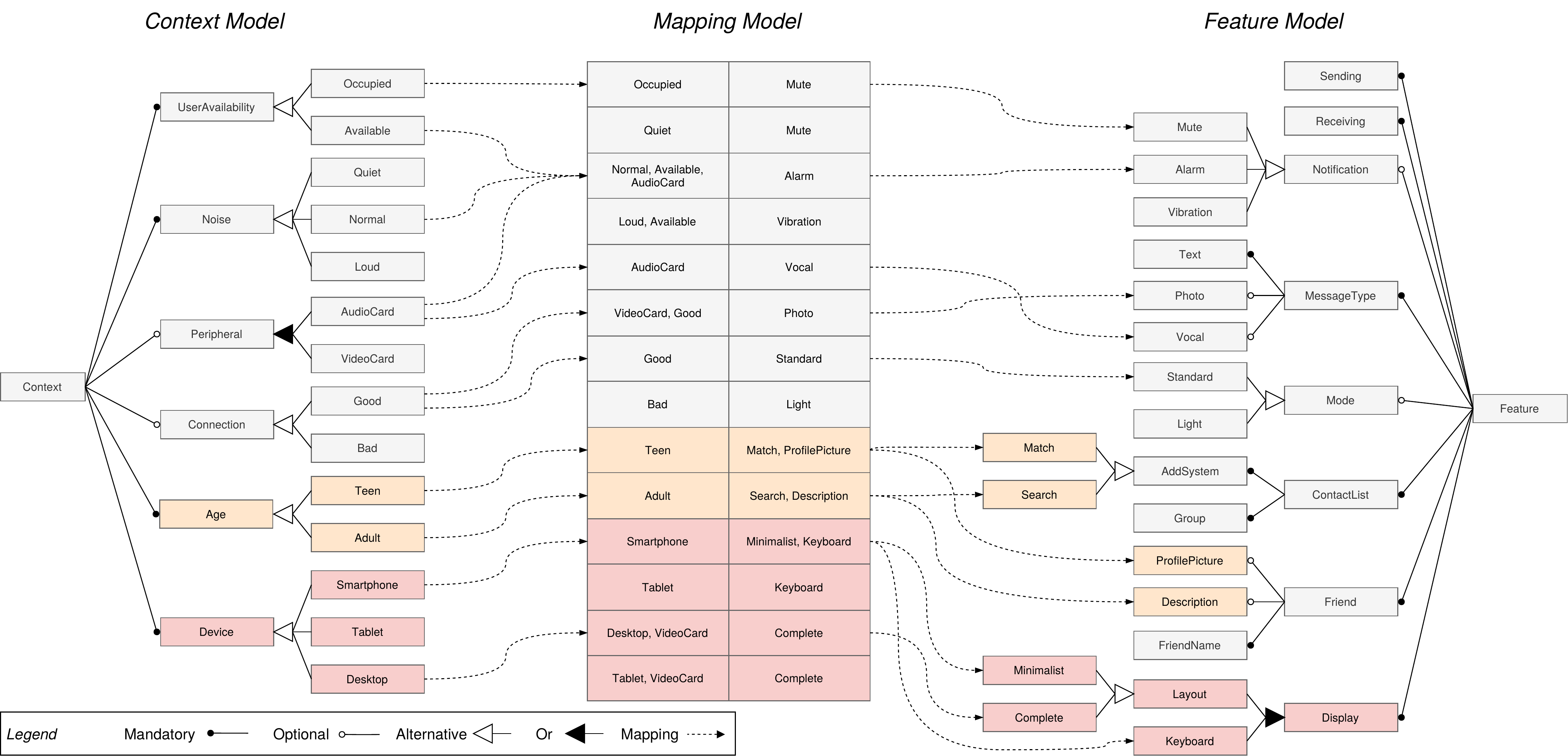}
  \caption{Design of a prototype of our smart messaging system. The context model is on the left, the feature model on the right and the mapping model between them in the middle. We voluntarily omitted some mapping details for readability reasons.}
  \label{fig:casestudy-models}
\end{figure*}

\subsection{Feature diagram}

Taking inspiration from Hartmann and Trew's \emph{multiple product line feature model}~\cite{Hartmann08FeaturesContexts},
in feature-based context-oriented programming, 
both the contexts and features of a FBCOP program are designed as feature diagrams.
%Now we described roughly our smart messenger system, we will introduce the notion of feature diagram.% with an example based on our case study.
%Such a feature diagram for our case study can be found on the right part of Figure~\ref{fig:casestudy-models}.
%
A feature diagram~\cite{Kang90FODA} describes the commonalities and variabilities of a system in terms of a tree-like structure where nodes represent features (or contexts) and edges the constraints between them. 
%A feature diagram describes the common functionalities of a system and its variants~\cite{Kang90FODA}. Such a diagram is a tree-like structure where the node are the features and the edges are the constraints between them. 
Examples of feature diagrams representing, respectively, the context and feature model of our smart messaging system, can be found on the left and right hand side of Figure~\ref{fig:casestudy-models}.

Constraints can either be hierarchical or cross-tree constraints.
One kind of hierarchical constraints are 
\emph{mandatory} constraints ensuring that a given child feature is always present if its parent feature is selected. Examples of such features in Figure~\ref{fig:casestudy-models} are \emph{Sending} or \emph{Receiving}. 
\emph{Optional} constraints express that a child may be selected if its parent is, as for the \emph{Photo} or \emph{Vocal} features.
\emph{Or} (resp. \emph{alternative}) constraints impose that at least (resp. exactly) one child should be selected if the parent is. An example of an \emph{or} constraint is the \emph{Display} feature: the messaging system can either display a \emph{Layout}, a \emph{Keyboard} or both to its users. %The kind of \emph{Notification} is an example of an \emph{alternative} constraint since our system may only propose one notification type when users receive new messages.
The kind of \emph{Layout} is an example of an \emph{alternative} constraint since the system will use either a \emph{Complete} or a \emph{Minimalist} layout but never both simultaneously.

Cross-tree constraints can be used to express exclusions or requirements between features.
An \emph{exclusion} constraint between features ensures only one feature is selected in the system at the same time.
A \emph{requirement} constraint between two features means the source feature can be selected only if the target feature is already selected.

A configuration of a feature diagram is a selection of features and is said to be valid if the selection satisfies all the constraints imposed by the diagram.

\subsection{FBCOP architecture}

\begin{figure}[!ht]
  \centering
  \includegraphics[draft=false, width=\linewidth]{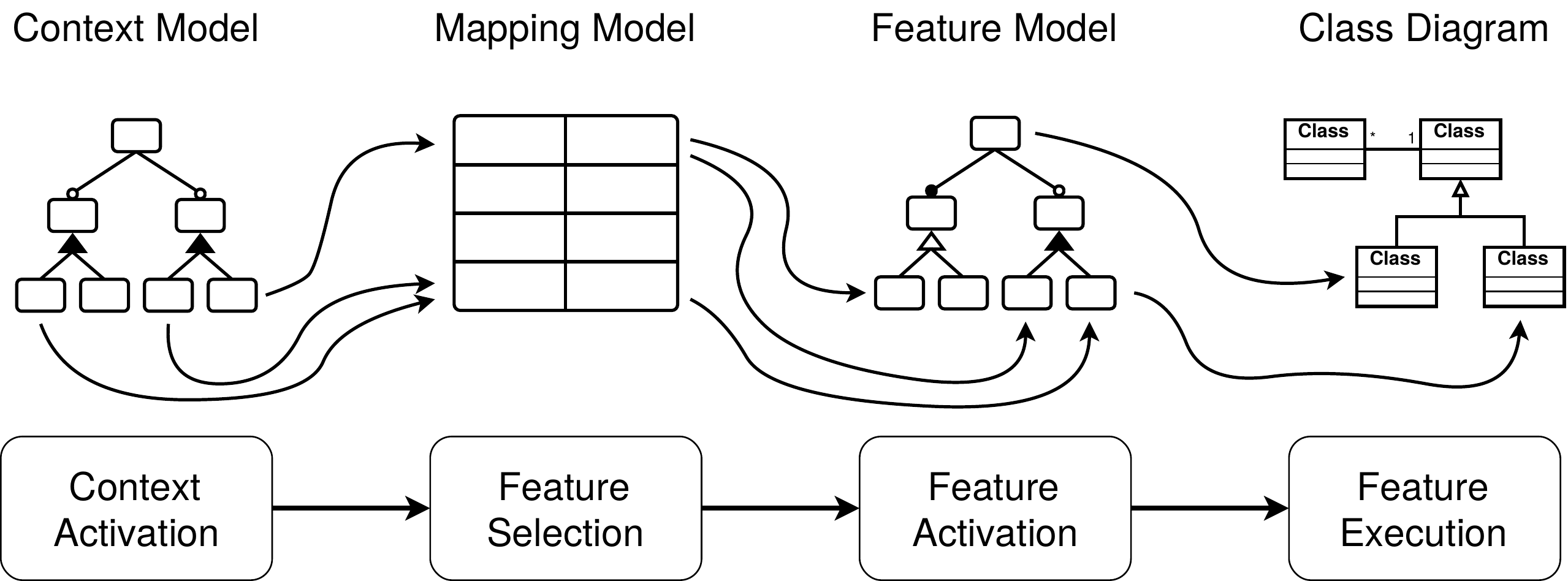}
  \caption{Overview of the different models and phases in a feature-based context-oriented system~\cite{Duhoux19ImplFBCOPL}.}
  \label{fig:FBCOApproach}
  %\vspace{-4mm}
\end{figure}

Figure~\ref{fig:FBCOApproach} sketches the overall FBCOP architecture. Using our running example we will
explain how this architecture enables a FBCOP system to adapt its behaviour to sensed context changes.
%Now we will explain how FBCOP adapts the behaviour of the system when context changes are sensed.
%Figure~\ref{fig:FBCOApproach} depicts the FBCOP architecture showing its overall lo\-gic and models involved.
For more information on how the FBCOP architecture is designed and implemented, please consult our paper~\cite{Duhoux19ImplFBCOPL}.

When a particular situation in the surrounding environment is (no longer) sensed, the system attempts to (de)activate it in its context model.
This happens during the \emph{context activation} phase.
For that, the system updates the current configuration of the context model, representing the current state of the surrounding environment, with the newly (de)activated context.
If the updated configuration still satisfies all constraints of the context model, this new configuration is kept and the system proceeds to the second \emph{feature selection} phase. Otherwise, the system rolls back to the previous valid configuration of the context model and ignores the new context.
Based on our case study, assume an \emph{Adult} uses the smart messaging system on her \emph{Smartphone}, is currently \emph{Available} for conversation, and that the ambient \emph{Noise} level is \emph{Normal}. As this configuration is valid, the system activates these contexts.

Next, the system determines what features are triggered by these new contexts through the mapping model.
This mapping model consists of a conjunction of individual mappings, each representing a \emph{N:N} relation from contexts to features.
In our example, the system would select the following features: \emph{Search}, \emph{Description}, \emph{Minimalist}, \emph{Keyboard}.
This selection then gets processed by the system in the \emph{feature activation} phase to attempt to activate (or deactivate) the corresponding features in its feature model.
Similar to the \emph{context activation} phase, the system updates the current configuration of the feature model, describing the currently active features, with the newly (de)activated features.
If the updated configuration is valid,  
%satisfies all constraints of the model, 
the system keeps this updated configuration. Otherwise it ignores this attempt.
In our example, as the new configuration of the feature model remains valid, the system activates this selection.

Finally, during the \emph{feature execution} phase,  the system adapts its behaviour by installing (resp. removing) the code associated to the selected features in the running code. % to refine or change its observable behaviour.
In our example, after installing these features, 
the user will be able to \emph{Search} for friends to add them to a group conversation, as well as to add a detailed \emph{Description} to her user profile. As she uses a \emph{Smartphone}, a \emph{Minimalist} layout and virtual \emph{Keyboard} will be presented by the system.

%% file: Testing.tex
\section{Testing}
\label{sec:case-testing}

Having introduced the notions of context-oriented programming and feature-based context-oriented programming, and illustrated them on our running example, let us revisit our three RQs in detail.

\subsection{Test suite generation}
Because of the large number of possible combinations of contexts and features, and the dynamicity of the contexts that trigger these features, exhaustively testing all possible scenarios induced by a such a program becomes intractable. The first research question (RQ1) is thus: \textbf{How to generate a pertinent yet tractable set of test scenarios for a given FBCOP application?}

Each such test scenario is a configuration of the context and feature models, representing a set of situations in the environment together with a set of features to activate or deactivate. 
It is worth noting that in the feature-based context-oriented approach considered, contexts (and features) are binary entities meaning that they are either active or not. This is the case even for contexts based on continuous measures such as the \emph{\textbf{Noise}} level. To distinguish between different noise levels, they are discretised into a partition of possible ranges, represented by child contexts such as \emph{\textbf{Quiet}}, \emph{\textbf{Normal}} and \emph{\textbf{Loud}} in an alternative relationship.
%Discretization has already been long studied before, especially in machine learning \cite{kotsiantis2006discretization}, which gives us a broad perspective of possible methods.
%
For simplicity's sake, let us assume that we have an %feature-based context-oriented 
application with these three contexts \emph{\textbf{Quiet}}, \emph{\textbf{Normal}} and \emph{\textbf{Loud}} and two features \emph{Photo} and \emph{Alarm}.

\begin{table}[!ht]
    \centering
    \caption{Example of a test suite, where \emph{A} (resp. \emph{D}) means that a  particular context or feature is \emph{A}ctivated (resp. \emph{D}eactivated).}
    \begin{tabular}{|c||c|c|c||c|c|}\hline
        Scenario & \textbf{Quiet} & \textbf{Normal} & \textbf{Loud} & Alarm & Photo \\\hline
        1 & D & A & D & A & A\\\hline
        2 & A & D & D & A & D\\\hline
        3 & D & A & A & D & D\\\hline
        \ldots & \multicolumn{5}{c|}{\ldots} \\\hline
    \end{tabular}
    
    \label{tab:example-test-suite}
\end{table}

Table \ref{tab:example-test-suite} lists a subset of the $2^5$ possible combinations of these contexts and features. For clarity, we keep the contexts on the left, the features on the right and highlight the contexts in bold.
%Note that 
Some of these scenarios may be invalid when they do not respect the constraints imposed by the context, feature or mapping model. Given the \emph{alternative} constraint in the context model between the contexts \emph{\textbf{Quiet}}, \emph{\textbf{Normal}} and \emph{\textbf{Loud}}, only one of them can be active at a time, making the third scenario invalid. The second scenario is invalid since the mapping does not permit the feature \emph{Alarm} to be active in context \emph{\textbf{Quiet}}. 

In Section \ref{sec:cit}, we will address RQ1 by reducing an FBCOP system to a highly reconfigurable system in presence of constraints, in order to follow a pairwise testing approach \cite{cohen2008constructing}. Such computation exploits a SAT-based representation unifying the context, mapping and feature models of an FBCOP system, effectively handling the three layers of constraints present in these systems. Moreover, this approach
can take advantage of particular  
%synergies with intrinsic 
properties of FBCOP models, as will be shown in Section \ref{sec:complexity} on complexity, and Section \ref{sec:optimization} which will highlight the high efficiency of two well-known optimizations on FBCOP systems.

To our knowledge, there is little research on this subject and focused on context-oriented programming. Whereas in this paper we mainly rely on Cohen's approach \cite{cohen2008constructing}, in the future we will further improve our approach by taking insights from pairwise testing in software product lines \cite{perrouin2012pairwise} and more diverse testing approaches \cite{do2012strategies}. Close fields include for example multi-objective \cite{henard2013multi} or many-objective test generation \cite{hierons2020many}. 

\subsection{Creation cost of a suite}
\label{sec:creation-cost}
The presentation of scenarios as shown in Table~\ref{tab:example-test-suite} quickly becomes unreadable, even for small systems with only a few tens of contexts and features. A solution to this problem is to show only 
%to restrict the list to 
the activation switches of contexts and features between subsequent scenarios instead. This idea is illustrated in Table~\ref{tab:example-test-suite-switches}.

\begin{table}[!ht]
    \caption{Example of a test suite with activation switches}
    \centering
    \begin{tabular}{|c|c|c|c|}\hline
        Scenario & Activation & Deactivation \\\hline
        1 & \textbf{Normal}, Alarm, Photo &  \\\hline
        2 & \textbf{Quiet} & \textbf{Normal}, Photo\\\hline
        3 & \textbf{Normal}, \textbf{Loud} & \textbf{Quiet}, Alarm \\\hline
    \end{tabular}
    \label{tab:example-test-suite-switches}
\end{table}

In addition to being more compact, 
this representation is closer to how an FBCOP system
works by dynamically activating and deactivating contexts and features.
This representation
tells a developer
exactly which context switches need to be simulated in the different test scenarios.
Limiting the number of context switches, which we will call the \emph{creation cost} of a test suite, is thus of fundamental importance if one wants to reduce simulation costs, which leads us to  
the second 
research question (RQ2): \textbf{How to minimise the effort of creating these generated test scenarios?}

First, we observe that the particular ordering of scenarios in a test suite may affect the number of context (de)activation switches.
E.g., the above example has 6 switches. By swapping scenario 1 and 2, the creation cost would get reduced to 4, as illustrated in Table \ref{tab:example-test-suite-switches-reordered}. 
In Section \ref{sec:test-rearrangement} %provides an answer to RQ2 
we will generalise this observation into a lightweight test suite rearrangement algorithm,
based on the creation cost, a metric tailored to FBCOP.
Other rearrangements were studied previously in software product lines, such as using statistical prioritization \cite{DPCSLSH17}, graph representation and heuristics minimising the analysis effort \cite{lity2017optimizing}, or similarity-based product prioritisation improving feature interaction coverage as fast as possible during testing \cite{al2019effective}.
\begin{table}[h]
    \caption{Rearranged version of the example test suite}
    \centering
    \begin{tabular}{|c|c|c|c|}\hline
        Scenario & Activations & Deactivations \\\hline
        2 & \textbf{Quiet}, Alarm &  \\\hline
        1 & \textbf{Normal}, Photo & \textbf{Quiet}\\\hline
        3 & \textbf{Loud} & Alarm, Photo \\\hline
    \end{tabular}
    \label{tab:example-test-suite-switches-reordered}
\end{table}

\subsection{System evolution}
\label{sec:evolution}
Finally, it is inevitable that developers will continue to evolve their system. This evolution involves the addition of novel features and contexts that would trigger the selection and activation of such features. For efficiency reasons and since existing test suites already consider all combinations of interest for the initial system design, this evolution could be done without recomputing the entire suite. This prompts the third research question (RQ3): \textbf{How to incrementally adapt a previously generated set of scenarios upon evolution of the application to a new version?}

Let us illustrate this situation by augmenting the example of Table \ref{tab:example-test-suite}. Assume that we would like to add a context \emph{\textbf{Tablet}} and corresponding feature \emph{Complete} to use a layout that makes full use of the tablet's screen. We extend the original table\footnote{Due to space limitations we replaced the original contexts \emph{\textbf{Quiet}} and \emph{\textbf{Normal}} and their assignments by ``\ldots''} by keeping the same scenarios and adding a column \emph{\textbf{Tablet}} and \emph{Complete} and then assigning them some activation choices (in italics) in Table \ref{tab:example-test-suite-augmented}.
%\begin{small}
\begin{table}[!ht]
    \caption{Example of an augmented test suite}
    \centering
    \begin{tabular}{|c||c|c|c||c|c|c|}\hline
        Scenario & \ldots & \textbf{Loud} & \textit{\textbf{Tablet}} & Alarm & Photo & \textit{Complete}\\\hline
        1 & \ldots & D & \textit{A} & A & A & \textit{D}\\\hline
        2 & \ldots & D & \textit{A} & A & D & \textit{A}\\\hline
        3 & \ldots & A & \textit{D} & D & D & \textit{D}\\\hline
        \ldots & \multicolumn{6}{c|}{\ldots} \\\hline
    \end{tabular}
    \label{tab:example-test-suite-augmented}
\end{table}
%\end{small}

Unfortunately, with these values, the first scenario is no longer valid. Indeed, if feature \emph{Photo} is active, then according to the mapping, context \emph{\textbf{VideoCard}} must be active too. Moreover, if both \emph{\textbf{VideoCard}} and \emph{\textbf{Tablet}} are active, then feature \emph{Complete} must be active too, 
which is inconsistent with its assigned activation choice \textit{D}.

The presence of constraints thus complicates the task of augmenting a previous test suite.
%As we can see, the presence of constraints makes the task of augmenting a previous test suite difficult. 
Augmentation by random assignment risks making test scenarios invalid, and doing it manually would be quite cumbersome.
%trying to do it manually is cumbersome, as is illustrated by our example.
What we need is an efficient and automatic way to assign an activation choice to the new contexts and features that keeps each test scenario valid. %Hence, an answer to RQ3 is discussed 
Section \ref{sec:incremental-testing} will present such an augmentation algorithm including two different update strategies.

%\authorcomment[comment]{PM}{No related work for the evolution part yet. None of the references from the review or that I found were about exactly this subject; "incrementality" and "evolution" in SPL refers to changing from a product to another because of, e.g., a context variation. It is not a static evolution (or FM expansion) as we do here.}

\subsection{Overview}

When programming FBCOP systems, developers need system testing in order to debug and evolve their system. The goal of our proposed testing methodology is to help them with the definition of interesting and evolving test suites.

\begin{figure}[!ht]
  \centering
  \includegraphics[draft = False, scale = 0.44]{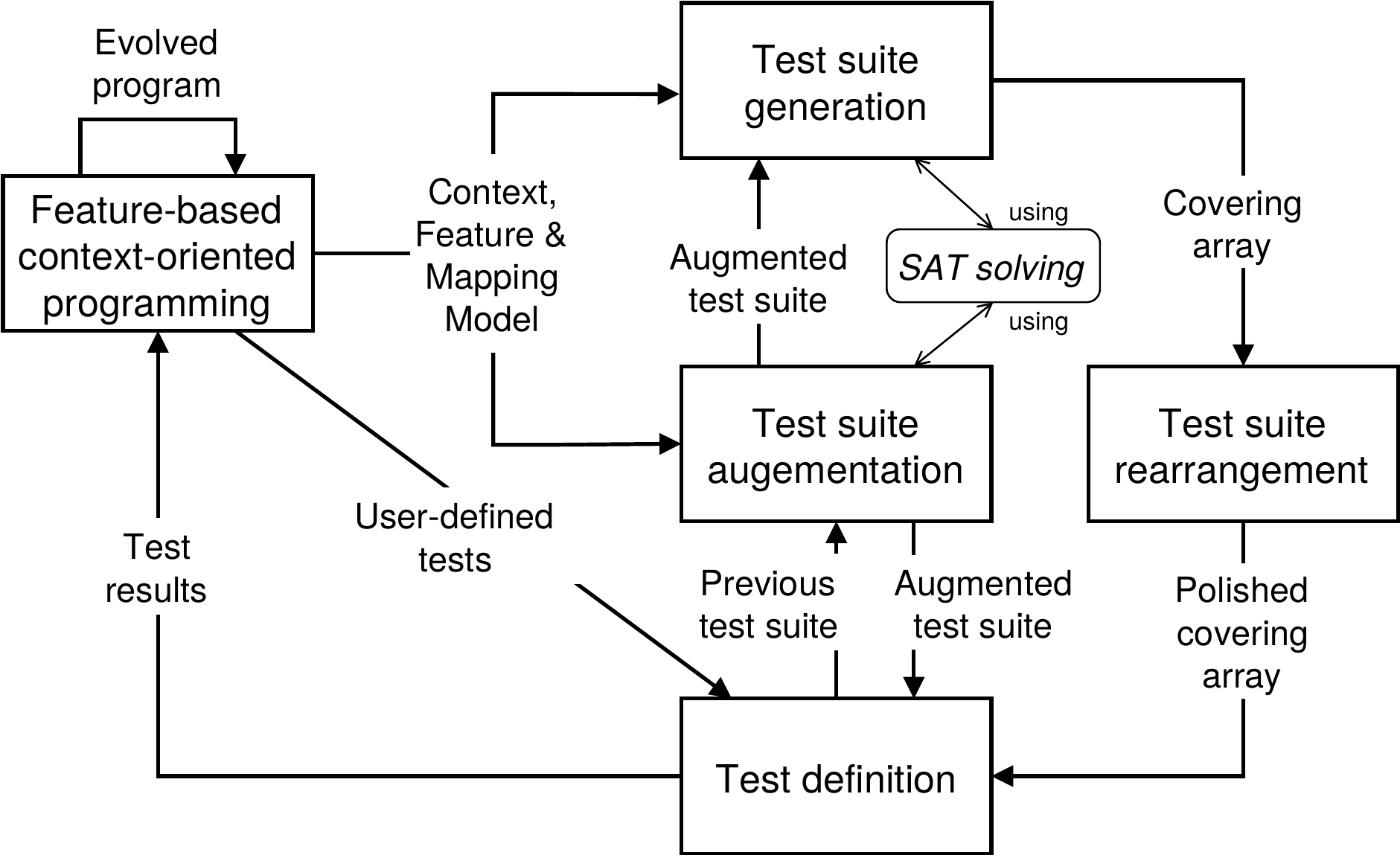}
  \caption{Summary of our FBCOP testing methodology}
  \label{fig:methodology-diagram}
  \vspace{-4mm}
\end{figure}

RQ1 will be addressed in Section \ref{sec:cit}. Our algorithms for rearranging a test suite (RQ2) and incrementally augmenting it (RQ3) are the subject of Sections \ref{sec:test-rearrangement}  and \ref{sec:incremental-testing}, respectively. Together, the proposed algorithms form a complete process schematized in Figure \ref{fig:methodology-diagram}.

%% file: CIT.tex
\section{Test suite generation}
\label{sec:cit}
To answer \textbf{RQ1}, we need a test scenario generation algorithm for FBCOP. Our answer consists in reducing a FBCOP system to a highly re-configurable system with constraints and to rely on Cohen's well-known \emph{combinatorial interaction testing} (CIT) algorithm to generate test suites for such systems \cite{cohen2008constructing}. This algorithm, which is based on a sampling methodology, takes as input a set of features with their variants (i.e., values that can be taken by a given feature) and a set of constraints on such variants expressed in \emph{conjunctive normal form} (CNF). It uses SAT solving to produce a test suite that covers any valid pair of variants. 

%Our test scenario generation algorithm (RQ1) relies on Cohen's well-known \emph{combinatorial interaction testing} (CIT) algorithm to generate test suites for highly re-configurable systems with constraints \cite{cohen2008constructing}. This algorithm, which is based on a sampling methodology, takes as input a set of features with their variants (i.e., values that can be taken by a given feature) and a set of constraints on such variants expressed in \emph{conjunctive normal form} (CNF). It uses SAT solving to produce a test suite that covers any valid pair of variants. 

% I see what you mean, I think "value" and "variants" are used correctly now
% a scenario contains "variants", but each features has a number of possible "values" INDEED ! THATS IT

% Assume a system with $k$ features $f_i$ each having $l_i$ possible values (i.e., variants) -> here I would say value : It is a value for a given
% feature and that value for that feature can give rise to a feature value pair (f,v) that is a variant
% I think it's clear in this sentence that variant=value ? BE PRECISE. Use variant when you mean an (f,v) pair use value when you just mean a value v
Assume a system with $k$ features $f_i$ each having $l_i$ possible values (i.e., variants). A pair of variants (($f_i$,$v_i$), ($f_j$,$v_j$)) is covered by a scenario when the specific value $v_i$ for $f_i$ (among the $l_i$ possible values) is in the scenario along with the specific value $v_j$ for $f_j$. 
Suppose an initially empty list of test scenarios $A$ and a list of \emph{uncovered} valid pairs of variants (e.g., by generating all pairs, then pruning the invalid ones through SAT solving) that are not yet covered by this list of scenarios (no test scenario contains both of these variants). The algorithm works by adding new scenarios until all pairs are covered.  A new scenario is selected by generating M candidate scenarios and selecting the one with highest coverage of uncovered pairs.\footnote{M is an input parameter to Cohen's algorithm which we experimentally assigned to 30 for our case study.} Each candidate scenario is constructed as follows:
\begin{enumerate}
    \item Find the value $v$ (among $l$ different values) and feature $f$ that is present in most uncovered pairs. 
    \item Let $f$ = $f_1$, then assign randomly $f_2, \ldots, f_k$ to each remaining feature.
    \item Assume all features $f_1$,\ldots,$f_{n-1}$ have been assigned to $v_1$, \ldots, $v_{n-1}$ 
    and that this partial assignment respects the constraints. 
    Now, find a value $v_n$ for $f_n$ (among $l_n$ possible values), such that a maximum of pairs 
    (($f_1$,$v_1$), ($f_n$,$v_n$)), \ldots, (($f_{n-1}$,$v_{n-1}$), ($f_n$,$v_n$)) 
    are still uncovered and that adding $v_n$ to the partial assignment keeps it valid (checked through SAT solving).
\end{enumerate}

%\noindent
%Observe that, due to the random choice of $M$ candidates, the CIT algorithm above is a sampling algorithm of which the results may change between two executions. The algorithm is meant to handle features with an arbitrary number of variants. In our context, we restrict to Boolean features representing two possible variants: the feature is activated or deactivated. Note that our results could be generalized by exploiting the extension presented in \cite{HLC13}.

\subsection{Covering Array Generation for FBCOP}

%As explained above, 
The CIT algorithm that we would like to use takes as input a set of constraints and a list of contexts and features that can be either activated or deactivated. In this section, we show how to reduce the problem of test suite generation for FBCOP systems to one of test suite generation of a highly re-configurable system with constraints to which the CIT algorithm applies. 

As described in Section \ref{sec:case} and depicted in Figure \ref{fig:casestudy-models}, a FBCOP system consists of a context model, mapping model and feature model. Both the context and feature models are represented as \textit{feature diagrams} (FD) representing a set of combinations of activated or deactivated (contexts or) features.  
In this representation, (contexts or) features can be modeled as Boolean variables whose assignments represent their activation status. E.g., if a Boolean variable representing the feature \emph{MessageType} is set to 1 (resp. 0), this means that this feature is activated (resp. deactivated). Any FD can be turned into conjunctive normal form \cite{czarnecki2007feature} whose Boolean atoms are the Boolean variables corresponding to the features' activation status. Any valid assignment of such a set of constraints represents the list of (de)activated (contexts or) features from the FD. 

Now that we know that the feature diagrams for both the context and feature model can be represented by constraints, the only thing left to do is to convert the mapping model into a set of constraints as well. %If this can be achieved, then our reduction will be complete.
The set of constraints will then be the conjunction of all constraints generated by each of the three models of a FBCOP system, and the list of contexts or features to activate or deactivate for a given test scenario will be modeled by Boolean variables. 

\begin{table}[h]
    \caption{Subset of the mapping model of our case study}
    \centering
    \begin{tabular}{|c|c|}\hline
        Occupied & Mute \\\hline
        Quiet & Mute \\\hline
        Normal, Available, AudioCard & Alarm \\\hline
        Loud, Available & Vibration \\\hline
    \end{tabular}
    \label{tab:small-mapping}
\end{table}
%\begin{figure}[!h]
%  \centering
%  \includegraphics[draft=false, scale=0.5]{images/small-mapping.png}
%  \caption{Subset of the mapping of our case study}
%  \Description{}
%  \label{fig:small-mapping}
%\end{figure}

Converting our mapping model to a set of constraints is straightforward. Remember that the mapping can be represented as set of N:N correspondences between contexts and features, as exemplified in Table \ref{tab:small-mapping}. A naive way to convert the mapping represented by this table to propositional logic would be to take each line as a simple equivalence relation between the left and right columns, 
%\begin{equation} Occupied \Leftrightarrow Mute \end{equation}
%\begin{equation} Quiet \Leftrightarrow Mute \end{equation}
%\begin{equation} Normal \land Available \land AudioCard \Leftrightarrow Alarm \end{equation}
%\begin{equation} Loud \land Available \Leftrightarrow Vibration \end{equation}
but this conversion is actually incorrect. Indeed, the equivalence between \emph{Occupied} and \emph{Mute} and the one between \emph{Quiet} and \emph{Mute} would imply the equivalence of the two contexts \emph{Occupied} and \emph{Quiet}. As contexts \emph{Quiet} and \emph{Loud} are mutually exclusive (they are part of an alternative constraint), context \emph{Occupied} would never be active at the same time as context \emph{Loud} under these constraints. 
In reality however, contexts \emph{Occupied} and \emph{Loud} can be active at the same time. 
We therefore propose a more liberal conversion where the activated contexts imply the activation of the corresponding features:

\begin{equation} Occupied \Rightarrow Mute \end{equation}
\begin{equation} Quiet \Rightarrow Mute \end{equation}
\begin{equation} Normal \land Available \land AudioCard \Rightarrow Alarm \end{equation}
\begin{equation} Loud \land Available \Rightarrow Vibration \end{equation}

In order to guarantee a correct mapping, one also needs to make sure that when features are activated, at least one combination of contexts for which that feature becomes active must be activated too. We do this by aggregating the different combinations of contexts that activate the same features :

\begin{equation}Mute \Rightarrow Occupied \lor Quiet  \end{equation}
\begin{equation}Alarm \Rightarrow Normal \land AudioCard \land Available\end{equation}
\begin{equation}Vibration \Rightarrow Loud \land Available \end{equation}

\noindent
From these two sets of Boolean formulas in propositional logic, we can obtain a set of CNF formulas, and thereby complete our conversion of a FBCOP system to a list of (contexts or) features with (Boolean) variants, in order to generate a covering array, and a set of CNF Boolean formulas, to be handled by a SAT solver. 
In Table \ref{fig:result-without-reordering}, we show three test scenarios of a generated test suite for the example presented in Section \ref{sec:case}. The whole test suite has 18 scenarios, and its creation cost (i.e., the number of context switches) is 161. Over 30 different runs of the CIT algorithm, the average size and cost of the generated scenarios are 17.36 and 151.9, respectively.

Interestingly, Scenario 13 in Table \ref{fig:result-without-reordering}, while correct, highlights an error in the design of our case study: it seems possible to have a \emph{Tablet} without an appropriate \emph{Layout}, when the \emph{VideoCard} context is not activated. It could be solved by adding a cross-tree constraint imposing that each \emph{Tablet} must have a \emph{VideoCard}. Finding such design errors is precisely why we need a good testing methodology.

\small
\begin{center}
\begin{table*}[!ht]
    \caption{Three test scenarios of a generated test suite}
    %\hspace*{-1cm}
    \begin{tabularx}{\textwidth}{|c|X|X|X|X|}
    \hline
    Scenario & Context activations & Context deactivations & Feature activations & Feature deactivations \\\hline
\ldots & \multicolumn{4}{c|}{\ldots}\\\hline

11 & Good, AudioCard, Tablet, Connection, VideoCard, Peripheral, Quiet & Normal, Smartphone & Notification, Complete, Standard, Vocal, Mode, Mute, Photo & Minimalist\\\hline
12 & Normal, Teen, Desktop & AudioCard, Tablet, Adult, Quiet & Match, ProfilePicture & Search, Description, Notification, Keyboard, Vocal, Mute\\\hline
13 & AudioCard, \textit{Tablet}, Quiet & Good, Normal, Connection, VideoCard, Desktop & Notification, Keyboard, Vocal, Mute & Complete, \textit{Layout}, Standard, Mode, Photo\\\hline

\ldots & \multicolumn{4}{c|}{\ldots}\\\hline
    \end{tabularx}
    \label{fig:result-without-reordering}
    %\vspace{-4mm}
\end{table*}
\end{center}
\normalsize

\subsection{Observations on Complexity}
\label{sec:complexity}
In the algorithm above,  
context and feature models are treated equally, in the sense that we take the conjunction of all constraints, independently of the model they represent. This means that the CIT algorithm will generate a covering array of strength 2 that covers all 2-way interactions between contexts and features alike (i.e., context-context, context-feature and feature-feature interactions). 

The reason for this choice is that taking a conjunction is more efficient than trying to combine subcovering arrays that would be obtained by applying the CIT algorithm to each context / mapping / feature model separately. For example, there could exist scenarios for the covering array of the feature model for which there would be no corresponding context model configuration. 
We would thus have to check any combination of test scenarios for the context model with scenarios from the feature model and remove those that are not compatible with the mapping model. 
This removal could cause a loss of coverage, forcing us to reuse the CIT algorithm in an iterative manner and with extra constraints until reaching a fixed point. This would be the same phenomenon as handling constraints after computing the array, which is known to be intractable~\cite{cohen2008constructing}. 

One could argue that taking into account the interaction between contexts and features is not logical as the testing should focus on the interactions between features only. However, because contexts are always directly related to features through the mapping, in practice we will generate almost no test scenarios whose goal will be to cover only pairs of contexts, or only pairs of features. Since the contexts are intrinsically related to the features through the mapping, if the models are well-designed, for a certain pair of contexts (or context-feature pair), another pair of features will always be present. E.g., the pair of context activations (\emph{Teen}: Activated, \emph{Good}: Activated) will always be tested with the combination of feature activations (\emph{Match}: Activated, \emph{Standard}: Activated), due to the mapping. If (\emph{Teen}: Activated, \emph{Good}: Activated) is an uncovered context pair, that also means that the feature pair is not covered either.

Another argument against our choice of not focusing on the interactions between features only, could be that adding contexts increases the number of tests. However, the size of the final covering array grows logarithmically in the number of contexts and features~\cite{cohen1997aetg}. Let $N_f$ be the number of features and $N_c$ the number of contexts. Assuming that both numbers are similar, we have: 

\begin{equation}
O\left(log\left(N_f \texttt{+} N_c\right)\right) \simeq O\left(log\left(2 N_f\right)\right)
\end{equation}
\begin{equation}
O\left(log\left(2 N_f\right)\right) = O\left(log\left(2\right) \texttt{+} log\left(N_f\right)\right)
= O\left(log\left(N_f\right)\right)
\end{equation}

\noindent Adding contexts thus does not affect
%Contexts can thus be added \textit{for free} without affecting 
the order of complexity of the algorithm if we assume that the number of contexts and features is similar (as is the case in our case study).

In addition to what has been said above, we believe that testing pairwise contexts is also important as it forces a developer to simulate each context and make sure that they can indeed be deployed. 

\subsection{Computation time and optimizations}
\label{sec:optimization}
Finally, we show two optimizations of CIT algorithms that work well with the constraint model induced by FBCOP:

\paragraph{1. Pre-processing of core and dead contexts and features.} 

There may be some contexts or features that always need to remain activated or deactivated in order to satisfy the constraints. It is known that identifying such core and dead feature reduces the effort from the SAT solver. Let C (resp. D) be the number of core (resp. dead) features and contexts, M the number of candidate scenarios, and S the size of a generated covering array. Identifying core and dead features allows us to save $\left(C \texttt{+} D\right) \times S \times M$ calls to the solver.

The simplest way to identify a dead (resp. core) feature is to check satisfiability of a partial configuration with only this feature being activated (resp. deactivated). If this partial configuration is not satisfiable, it means that the feature must always be deactivated (resp. activated) in all configurations.
% MENTIONED IN FUTURE WORK, SHALL WE JUST DELETE THIS ?
%It can be done in different ways, such as by heuristics exploiting the SAT solver \cite{johansen2012algorithm}. 
%or with \cite{}. %Need to find citation on model analysis

\paragraph{2. Additional step based on propagation}
An important step of SAT solvers is Boolean constraint propagation (BCP) \cite{marques1999grasp}. From a CNF formula and partial truth assignment (or partial configuration), the BCP phase has the objective of producing a \emph{unit clause}, that is a clause that contains a single unbound variable. The value of that unbound variable can then be inferred, as all clauses of a CNF formula must be True for the formula to hold.

To illustrate the intuition behind this propagation, let us consider contexts \emph{Quiet}, \emph{Normal} and \emph{Loud}, and the features \emph{Mute}, \emph{Alarm} and \emph{Vibration}. %As said above, 
We view these contexts and features as Boolean variables which are True (activated) or False (deactivated). Assume that context \emph{Quiet} has been assigned the value True. This context cannot be active if either contexts \emph{Normal} or \emph{Loud} are active. Hence, we can infer to assign the False value to those two contexts. We also know that context \emph{Quiet} implies feature \emph{Mute} via the mapping, and that \emph{Mute} cannot be active if either feature \emph{Alarm} or \emph{Vibration} are active. Consequently, we can then infer to assign a True value to \emph{Mute}, and False to \emph{Alarm} and \emph{Vibration}.

Cohen et al.~\cite{cohen2008constructing} proposed an optimization which exploits BCP. Let ${inf}_{V_1}$, \ldots, ${inf}_{V_m}$ be the $m$ values inferred by BCP, and ${inf}_{F_k}$ denote the features whose value ${inf}_{V_k}$ were inferred. The modification to step (3) of the CIT algorithm is as follows:

\begin{enumerate}
    \setcounter{enumi}{2}

    % REWRITTEN BY KM
    \item Assume all features $f_1$,\ldots,$f_{n-1}$ have been assigned to $v_1$, \ldots, $v_{n-1}$ 
    and that this partial assignment respects the constraints. 
    \textbf{If $f_n$ was already assigned an inferred value, skip this step. Otherwise,}
    find a value $v_n$ for $f_n$ (among $l_n$ possible values), such that a maximum of pairs 
    (($f_1$,$v_1$), ($f_n$,$v_n$)), \ldots, (($f_{n-1}$,$v_{n-1}$), ($f_n$,$v_n$)) 
    are still uncovered and that adding $v_n$ to the partial assignment keeps it valid (checked through SAT solving).
    % OLD VERSION
    %\item Assume all features between $f_1$ and $f_{n-1}$ have been assigned to $v_1, v_2, \ldots v_{n-1}$, and that this partial assignment respects the constraints. \textbf{If $f_n$ was already assigned a value, skip this step.
    %Otherwise,} find and choose a value $v_n$ for $f_n$ among the $l_n$ possibilities, such that a maximum of pairs $\left(v_1, v_n\right),$ ... $\left(v_{n-1}, v_n\right)$ are still uncovered and that adding $v_n$ to the partial assignment keeps it valid (checked through SAT solving).
    
    % REWRITTEN BY KM
    \item After choosing the value $v_n$ for feature $f_n$, find the propagated values ${inf}_{V_1}$, \ldots ${inf}_{V_m}$ thanks to SAT solving and assign to each feature ${inf}_{F_k}$ its corresponding value ${inf}_{V_k}$.
    % OLD VERSION
    %\item After choosing the value $v_n$ for feature $f_n$, find the propagated values ${inf}_{V_1}, {inf}_{V_2}, ... {inf}_{V_n}$ thanks to SAT solving and assign to each feature ${inf}_{F_k}$ its corresponding value ${inf}_{V_k}$.
\end{enumerate}
%\vskip{-1cm}
\noindent

Table \ref{tab:first-vs-second-opti} shows the number of propagations and the number of propagated values with or without the first optimization, when applied to our case study. It shows that these optimizations affect two largely separate aspects. Indeed, the number of propagated values is almost the same with or without the first optimization. This is mainly due to the design phase as it is unlikely that a core (resp. dead) context/feature which is by definition always activated (resp. deactivated) would have an impact on other highly dynamic features/contexts of a FBCOP system. %BCP works particularly well with highly-constrained systems that have high potential for propagation, which is the case of FBCOP.

\begin{table}[!ht]
    \caption{Impact of the pre-processing optimization on the number of propagations and propagated values.}
    \centering
    \begin{tabular}{|c|c|c|} \hline
         & Propagations & Propagated values \\\hline
        Without Optimization 1 & 12567 & 14282 \\\hline
        With Optimization 1 & 3327 & 13965 \\\hline
    \end{tabular}
    
    \label{tab:first-vs-second-opti}
    %\vspace{-6mm}
\end{table}

\begin{table}[h]
    \caption{Impact of the optimizations on the computation time}
    \centering
    \begin{tabular}{|c|c|c|}\hline
         & Computation time (seconds) \\\hline
        Without optimization & 17.2\\\hline
        Optimization 1 & 4.6 \\\hline
        Optimization 2 & 14.6 \\\hline
        Both optimizations & \textbf{2}\\\hline
    \end{tabular}
    \label{tab:result-computation-time}
\end{table}

%\noindent
Table \ref{tab:result-computation-time} illustrates how each of these optimizations improve the average computation time for 30 executions of the CIT algorithm on our case study. The pre-processing of core/dead contexts and features drastically prunes the set of features and corresponding constraints. The first optimization reduces the computation time by 73\%, the second one decreases it by (only) 15\%. The two optimizations combined reduce the time of computation by an average of 88\%.

%With this first optimization, the computation time gets reduced by 73\%. With the modified CIT algorithm based on BCP, a net decrease of computation time of (only) 15\% is reached. The two optimizations combined reduce the time of computation by an average of 88\%. %Observe that as the number of propagated values is approximately the same with or without preprocessing, we always save 2.5 seconds with the second optimization. 

To answer \textbf{RQ1}, we adapted a FBCOP system and proposed a test scenario generation algorithm. Its computation time is 2 seconds when applied to our case study, which highlights its lightweight nature.

%% file: Rearrangement.tex
\section{Test suite rearrangement}
\label{sec:test-rearrangement}

As discussed in Section \ref{sec:creation-cost}, by decreasing the creation cost the effort of creating a generated test suite can be minimised.
%Decreasing the creation cost also decreases the effort of using a generated test suite (RQ2), as discussed in Section \ref{sec:creation-cost}.
The creation cost of a test suite was defined as the number of context switches (i.e., activations or deactivations) needed to simulate all scenarios from the suite. In this Section, we answer \textbf{RQ2} by proposing a test suite rearrangement algorithm which minimizes this cost. 

We first define the \textit{distance} between two test scenarios to be the number of activated (resp. deactivated) contexts in the first scenario which get deactivated (resp. activated) in the second one. Thus, the distance between two test scenarios is the number of context switches needed to go from one configuration of the contexts to the other. The purpose of the rearrangement algorithm is to rearrange the different scenarios in such a way that the total distance between subsequent scenarios in the test suite remains as small as possible.

\begin{algorithm}[!ht]
\SetAlgoLined
\DontPrintSemicolon
%\KwResult{}
\SetKwInOut{Input}{input}
\SetKwInOut{Output}{output}
\Input{$L_0, t_0$}
\Output{$L$}
\Indp\Indpp
    $L$ = new List()\;
    test = $t_0$ \;
    \While{$L_0$ is not empty}{
        bestTest = minDistance($L_0$, test)\;
        $L$.add(bestTest)\;
        $L_0$.remove(bestTest) \;
        test = bestTest \;
    }
    return $L$ \;
 \Indp\Indpp
 \caption{Rearrangement algorithm}
 \label{alg:reordering-pseudo-code}
\end{algorithm}

The pseudo code of our rearrangement algorithm can be found in Algorithm \ref{alg:reordering-pseudo-code}. It takes a list of test scenarios \emph{$L_0$} and a default configuration \emph{$t_0$} where all contexts are deactivated which will act as initial scenario. The algorithm maintains an ordered list \emph{L} that is used to build the new rearranged test suite. The algorithm uses an auxiliary method $minDistance\left(L_0, test\right)$, whose goal is to find the test scenario in $L_0$ that is closest to a given scenario $test$.

As an example, for the test suite shown in Table~\ref{fig:re-ordered-result}, the rearranged test suite has a creation cost of 89 instead of 161 as for the original test suite which is a decrease of 45\%. Over 30 executions of the CIT algorithm on our case study, the creation cost of the rearranged test suites is 85, as opposed to 152 for the original test suites. We observed a stable decrease of 44\% in cost, with a variance of 4.5\% over all executions of the CIT algorithm.

\small
\begin{center}
\begin{table}[!h]
    \caption{Three test scenarios of a re-arranged test suite}
    %\hspace*{-1cm}
    \begin{tabularx}{0.5\textwidth}{|c|X|X|X|X|}
    \hline
    Scenario   & Context     & Context       & Feature     & Feature \\
               & activations & deactivations & activations & deactivations \\\hline
\ldots & \multicolumn{4}{c|}{\ldots}\\\hline
3 & Connection, Normal, Bad & Quiet & Mode, Light & Notification, Mute\\\hline
4 & Peripheral, AudioCard &  & Notification, Alarm, Vocal & \\\hline
5 & Smartphone, Teen & Adult, Tablet & Match, Layout, ProfilePicture, Minimalist & Description, Search\\\hline
\ldots & \multicolumn{4}{c|}{\ldots}\\\hline
    \end{tabularx}
    \label{fig:re-ordered-result}
    %\vspace{-6mm}
\end{table}
\end{center}
\normalsize

When applying a series of tests, it is important to isolate the features or interactions that are responsible for the failures. When a failure occurs in a given test scenario, it is highly likely that the failure is caused by an interaction with a new feature added in this test scenario. The goal of the rearrangement is to reduce the number of context switches. An indirect consequence is that it also drastically reduces the number of feature switches and hence the number of feature interactions to inspect. 

Our answer to \textbf{RQ2} is thus a simple-to-implement rearrangement algorithm which greedily tries to minimize the creation cost of a given test suite, a metric closely related to the effort of simulating and creating this test suite.

%% file: Incremental.tex
\section{Test suite augmentation}
%\section{Incremental test suite augmentation}
%\section{Incremental testing}
\label{sec:incremental-testing}

Like any software system, FBCOP systems are subject to continuous development that eventually expands the initial system's range of possible behaviours and contexts. Those additions, which introduce new contexts, features and relationships in the system's models, call for an augmentation of the test suite. For reasons of efficiency and stability, it is worth trying to achieve this augmentation without recomputing the entire test suite, as was discussed in Section \ref{sec:evolution}. Our answer to \textbf{RQ3} is then a greedy algorithm for incremental test suite augmentation that works in two steps:

%\begin{enumerate}
%    \item 
\paragraph{1.} The algorithm starts by updating existing test scenarios with variables representing new features/contexts that were added. The idea is to try to update each scenario from the original test suite. As each variable can take two values, the addition of $n$ features/contexts generates $2^n$ configurations from the original scenario, that is one for each possible Boolean assignment. If one such configuration remains valid, then we not only keep the pairs from the original test, but we also add pairs that cover the new features/contexts. Note that, since new features/contexts lead to new constraints, this may lead to test scenarios where none of the assignments remain valid. That is, some scenarios from the initial suite may not be expandable and will be dismissed. 
%    \item
\paragraph{2.} As second step, the algorithm checks if the augmented test suite is complete (i.e., complete coverage of the possible pairs). If not, the algorithm pre-processes the CIT algorithm with the current version of the augmented test suite, i.e. the initial list of test scenarios $A$ of the CIT algorithm is no longer empty. The generated test scenarios are then rearranged with the algorithm of Section \ref{sec:test-rearrangement} (updated test scenarios naturally keep their original order). The result is a complete augmented test suite. %This algorithm needs an input test scenario $t_0$ which we define to be the last test of the original test suite, in order to minimize the cost of re-attaching the generated test cases to the original test suite.
%\end{enumerate}

\subsection{Strategies for Step 1}
In practice, we do not want to enumerate all $2^n$ configurations in Step 1. Instead, we propose two strategies to generate a valid configuration for the original test scenario, if this is possible.

\paragraph{Strategy 1: Test update based on SAT solving} We update each scenario from the original suite with Boolean variables associated to the new features/contexts. We can feed this new partially assigned configuration to a SAT solver to try and generate a configuration that satisfies the constraints induced by the new system. The SAT solver will find assignments for the new Boolean variables such that the constraints are satisfied, hence leading to a new valid scenario, if there is one. Non-satisfiable test scenarios are dismissed.

\paragraph{Strategy 2: Test expansion based on features}
By default, incremental SAT solving will often assign the new contexts/features to False, or at least generate very similar configurations. That limits the coverage of the new pairs induced by the introduction of new contexts/features, as we do not inject much diversity in the old test suite.
We therefore propose to help the SAT solver through the random generation of ``partial scenarios''. These are partial configurations for which only the variables of the new features/contexts are assigned a value and variables representing features/contexts of the original system are left free. We try to randomize the generation of such partial scenarios so that they uniformly cover the set of new pairs. Such partial configurations are then combined with existing scenarios taken from the original test suite. Of course, due to model constraints, this may lead to invalid configurations. However, as opposed to the problem identified by Cohen et al.~\cite{cohen2008constructing}, we are not adding constraints after computing the whole array, but just checking compatibility for a restricted number of pairs.

Assume a test suite containing test scenarios $t_1$, \ldots , $t_l$, a LIFO queue $Q$ which contains partial scenarios $s_1$, \ldots $s_m$, and a maximum number of steps $S$. We now present an update procedure. We iterate over the following four steps until all test scenarios have either been updated or deleted :

\begin{enumerate}
    \item %Step 1: 
    Let $t_k$ be the current candidate test scenario to be updated,
    with $k\in \{1 \ldots l\}$. Pop $Q$ to retrieve partial scenario $s_{curr}$.
    \item %Step 2:
    Try to combine $s_{curr}$ with the next $S$ test scenarios $t_k$, \ldots, $t_{k\texttt{+}S \texttt{-} 1}$ consecutively. Stop if one test scenario $t_{k\texttt{+}i}$ is not compatible with partial scenario $s_{curr}$ (checked through SAT solving). 
    \item %Step 3: 
    Push $s_{curr}$ to $Q$.
    \item %Step 4: 
    If scenario $t_k$ has not been updated after $m$ iterations, then this means that we tested all combinations between $s_1, s_2, ... s_m$ and $t_k$. In such case, apply Strategy 1. 
\end{enumerate}

\noindent
The above algorithm selects $t_k$, the next test scenario to be updated and a partial scenario $s_{curr}$ (1). It then tries to combine $s_{curr}$ with the next $S$ scenarios (2) in order to identify and update those tests in $t_k$ \ldots $t_{k\texttt{+}S \texttt{-} 1}$ that are compatible with this partial scenario. All compatible scenarios are thus considered updated with the values of partial scenario $s_{curr}$, while the incompatible test scenario (if there is one) is left unchanged. Note that 
the bigger $S$ is, the less diversity there will be if a lot of scenarios are compatible with $s_{curr}$. On the other hand, keeping the same partial scenario reduces the number of modifications needed. The algorithm then selects the next partial scenario (3). In case a scenario $t_k$ cannot be updated by any partial scenario, we re-apply Strategy 1 on this test scenario (4).

\subsection{Performance}
\label{incrementality-performance}

We now evaluate the performance of the above strategies on our case study of Figure \ref{fig:casestudy-models}. We start with the grey contexts, mapping and features that we consider to be the original system. We then assume the system evolves into a second version that can adapt to a user's age, corresponding to the orange part of Figure \ref{fig:casestudy-models}. It comprises contexts \emph{Age}, \emph{Teen} and \emph{Adult} and features \emph{Match}, \emph{Search}, \emph{ProfilePicture} and \emph{Description}. 
After that, the system evolves again to include a set of features and contexts to obtain a responsive user interface, corresponding to the red part of Figure \ref{fig:casestudy-models}. It contains the features \emph{Display}, \emph{Keyboard}, \emph{Layout}, \emph{Minimalist} and \emph{Complete} and the contexts \emph{Device}, \emph{Smartphone}, \emph{Tablet} and \emph{Desktop}.

We define new metrics to study the performance of the augmentation procedure. Assume the original test suite contains $l$ tests. Consider an augmented test suite, containing both updated test scenarios and new generated test scenarios produced by the augmentation procedure. We define the \emph{modification cost} to be the number of new context switches introduced in step (1). We define the \emph{generation cost} to be the number of context switches introduced in step (2) (creation cost of the new test scenarios). The \emph{total cost} is their sum.

\begin{figure}[h]
\centering
\includegraphics[draft = False, width=.23\textwidth]{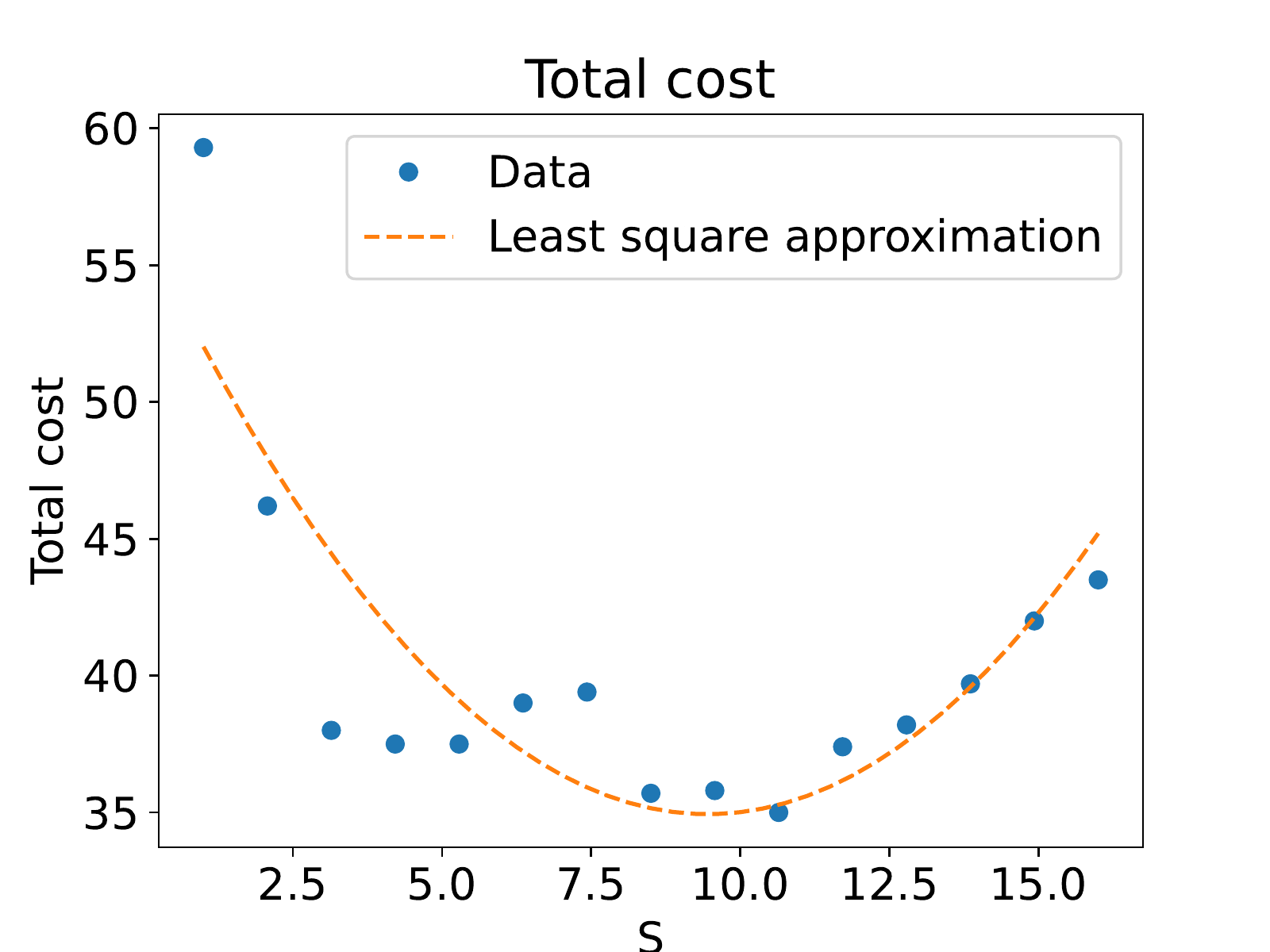}\hfill
\includegraphics[draft = False, width=.23\textwidth]{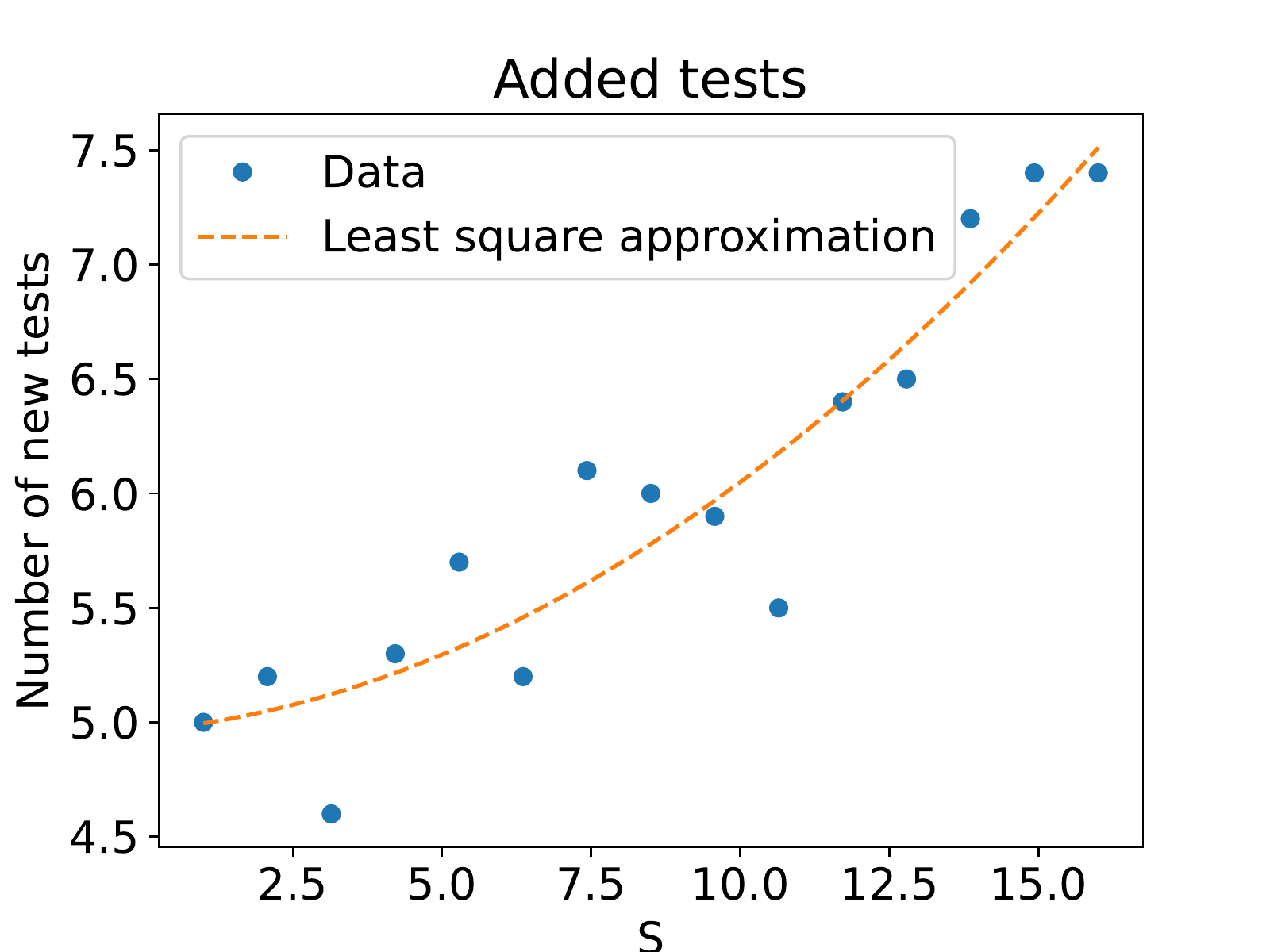}\hfill
\caption{Influence of parameter $S$ on the total cost of augmenting a test suite}
\label{fig:cost-S}
\end{figure}

We conduct two sets of experiments. In the first one, we explore the impact of the choice of the value of $S$ on Strategy 2. In Figure \ref{fig:cost-S}, we observe the expected compromise: small values of $S$ produce few new test cases (thus low generation cost) but high total cost due to high modification cost. Conversely, high values of $S$ produce many new tests, resulting in high total cost. A compromise can be around the value 9, with a total cost of 35.

\begin{figure}[h]
\centering
\includegraphics[draft = False, width=.23\textwidth]{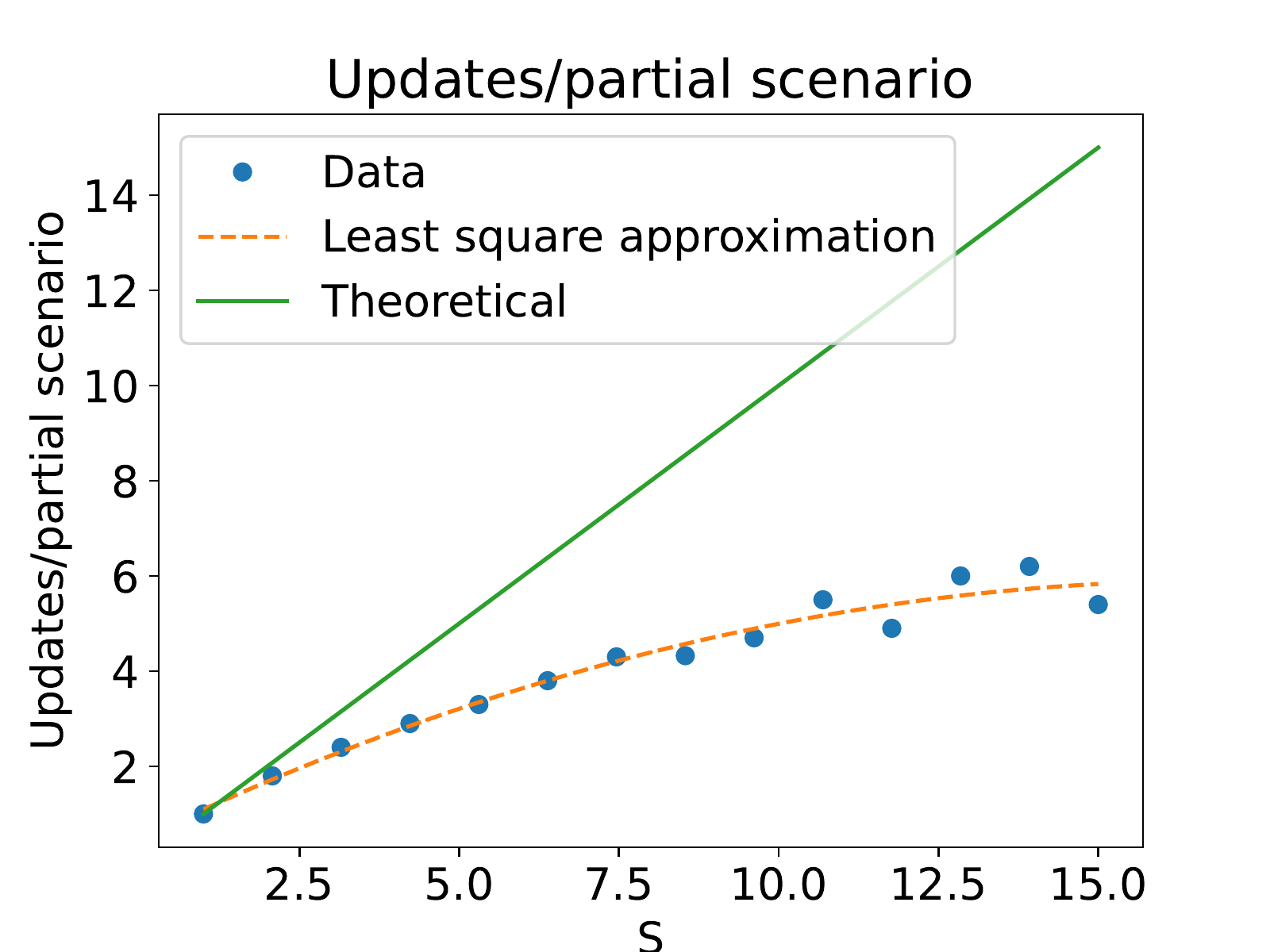}
\caption{Updates/partial scenarios, on average}
\label{fig:scenario-S}
\end{figure}

In Step 2 of Strategy 2, we combine partial scenarios to a maximum of $S$ consecutive test scenarios to update them while minimizing change. We study the actual number of consecutive test scenarios updated with a single partial scenario, on average, or \textit{updates/partial scenario}. In theory, this number should approach $S$. Instead, in Figure \ref{fig:scenario-S} we observe that this ratio flattens as S increases. Indeed, we have to change the partial scenario used if one incompatibility is found in this Step 2, and the probability of this phenomenon occurring increases with $S$. Hence, these incompatibilities force the algorithm to use different partial scenarios, no matter how $S$ increases. Observe also that high values of $S$ still tend to decrease the overall diversity of the updated scenarios, as illustrated with the growing number of new test cases needed in Figure \ref{fig:cost-S}.

%Assume partial scenarios $s_1, s_2, ... s_k$ were used to update test scenarios ($k$ can be lower than the total number of existing partial scenarios $m$), then we study how many times a partial scenario is used to update a test scenario, or \textit{updates/partial scenario}. In theory, this number should approach $S$. Instead, in Figure \ref{fig:scenario-S} we observe that this ratio flattens as S increases. Indeed, we have to change the partial scenario used if one incompatibility is found in this step 2, and the probability of this phenomenon occurring increases with $S$. Hence, these incompatibilities force the algorithm to use different partial scenarios, no matter how $S$ increases. Observe also that high values of $S$ still tend to decrease the overall diversity of the updated scenarios, as illustrated with the growing number of new test cases needed in Figure \ref{fig:cost-S}.

We now compare both incremental augmentation strategies. Table \ref{tab:system1-system2} shows the costs of augmenting the test suite when moving from the original to the second version of the system. Table \ref{tab:system2-system3} shows the results for the augmented test suite obtained when moving from the second to the third version. We compare those results with re-applying the CIT algorithm without taking any existing test suites into account (\emph{NOREUSE} row). 

\begin{table}[h]
    \caption{Comparison when going from version 1 to version 2. Best cost and size are highlighted in bold.}
    \centering
    \begin{tabular}{|c|c|c|c|}\hline
        Strategy & Modif. Cost & Total Cost & Size \\\hline
        NOREUSE & / & 113 & \textbf{15.7} \\\hline
        Strategy 1 & 2 & 22.5 & 20.3 \\\hline
        Strategy 2 & 8.33 & \textbf{11} & 15.9 \\\hline
    \end{tabular}
    \label{tab:system1-system2}
    %\vspace{-6mm}
\end{table}
\begin{table}[h]
    \caption{Comparison when going from version 2 to version 3. Best cost and size are highlighted in bold.}
    \centering
    \begin{tabular}{|c|c|c|c|}\hline
        Strategy & Modif. Cost & Total Cost & Size \\\hline
        NOREUSE & / & 152 & \textbf{17.4}\\\hline
        Strategy 1 & 4.9 & 46 & 27.4 \\\hline
        Strategy 2 & 5.2 & \textbf{35} & 21.4 \\\hline
    \end{tabular}
    \label{tab:system2-system3}
\end{table}

\noindent
\emph{NOREUSE} produces the least amount of test scenarios, which is expected since CIT's objective is to minimize such number. However, it can be very expensive as the CIT algorithm has to redo the entire work for each incremental evolution. As expected, Strategy 1 injects little diversity in the existing test scenarios (as illustrated by the low modification cost) but adds many more test scenarios. Strategy 2 addresses this problem and achieves the lowest total cost, reducing in the case of Table  \ref{tab:system2-system3} by more than a factor 4 the cost of creating a new test suite by reusing the previous one.

%% file: Future-work.tex
\section{Conclusion and Future Work}
\label{sec:conclusion}

This paper addressed three main research questions, namely:  \textbf{(RQ1)} How to generate a pertinent yet tractable set of test scenarios for a given FBCOP application? \textbf{(RQ2)} How to minimise the effort of creating these generated test scenarios? \textbf{(RQ3)} How to incrementally adapt a previously generated set of scenarios upon evolution of the application to a new version?

Our answer to RQ1 was to reduce the problem to one of computing test suites for highly re-configurable systems with constraints and pairwise testing. RQ2 was addressed by rearranging the test scenarios in such a test suite to minimize the number of context switches needed to simulate the entire suite. Finally, to answer RQ3 we explored an algorithm to incrementally augment existing test suites upon program evolution.

Whereas the focus of this paper was on test scenario generation for feature-based context-oriented programs, we believe the approach could be generalised to other context-aware systems if we obtain a set of contexts and features from them and discretize these contexts and features \cite{kotsiantis2006discretization}.

Validation of context-oriented programs is still at its infancy. There exist a wide range of possibilities to improve upon the work presented in this paper. We could improve Cohen's algorithm~\cite{cohen2008constructing} by exploiting the fact that our SAT formulas are mostly produced from feature diagrams in FBCOP. Exploiting such representation with algorithms such as those presented by Johansen~\cite{johansen2012algorithm} and Kowal~\cite{kowal2016explaining} could help identify dead/core features or other anomalies that should be embedded in or rejected from each test scenario. Another way would be to use Quantified Boolean Formula solvers~\cite{mauro2021anomaly}. Recent SAT solvers could compute randomly valid configurations of SAT formulas and have been shown to be efficient to estimate the t-wise coverage of various large size examples~\cite{PAPDC19,BLM20} but have not yet been adapted to handle systems with constraints.

%Recent SAT solvers could allow us to randomly compute valid configurations of SAT formulas. They have been shown to be efficient to estimate the t-wise coverage of various large size examples~\cite{PAPDC19,BLM20} but have not yet been adapted to handle systems with constraints.

The rearrangement process could also be extended by considering other objectives. In the spirit of Devroey et al.~\cite{DPCSLSH17}, one could target rearrangements that prioritize contexts that have been identified to be used frequently in deployed systems. We could also explore strategies that give more importance to some pre-defined high-frequency switches between contexts. Similarly, new ways to generate partial scenarios in the augmentation procedure could be considered.
%Another possible extension is to consider new ways to generate the partial scenarios used in the augmentation procedure. 

The main objective of this paper being to pose the foundations of our new theory, the experimental section was reduced to its minimum. So far, our prototype has been applied to a small academic case study only. The cost of creating a test suite for it and augmenting it has been reduced by more than half. In future work, we hope to investigate how its performance scales to industrial-scale case studies. 

%% file: Acknowledgements.tex
\section*{Acknowledgements}

Axel Legay and Pierre Martou are supported by the CYRUS project ---  Convention n° 8227, Pôle LiW --- Appel 27.